\documentclass[conference]{IEEEtran}
\IEEEoverridecommandlockouts

\usepackage{setspace}
\usepackage{cite}
\usepackage{amsmath,amssymb,amsfonts}
\usepackage{algorithmic}
\usepackage{graphicx}
\usepackage{textcomp}
\usepackage{xcolor}
\usepackage[hyphens]{url}

\def\BibTeX{{\rm B\kern-.05em{\sc i\kern-.025em b}\kern-.08em
    T\kern-.1667em\lower.7ex\hbox{E}\kern-.125emX}}

\usepackage{pifont}
\newcommand{\ncirc}[1]{\ding{\numexpr181 + #1\relax}}
\newcommand{\pcirc}[1]{\ding{\numexpr171 + #1\relax}}
\usepackage{amsmath}
\usepackage{svg}
\usepackage{graphicx}
\usepackage{microtype}
\usepackage{graphicx}
\usepackage{subfigure}
\usepackage{booktabs} 
\usepackage[dvipsnames]{xcolor}
\usepackage[normalem]{ulem}
\usepackage{booktabs}   
\usepackage{threeparttable} 
\usepackage{tabularx} 
\usepackage{adjustbox}
\usepackage{array}

\usepackage{graphicx}     
\usepackage{booktabs}     
\usepackage{array}        
\usepackage{wasysym}      
\usepackage{amssymb}      
\newcommand{\vhead}[1]{\rotatebox{90}{\parbox{2.2cm}{\centering\small\bfseries #1}}}
\newcommand{\ck}{$\checkmark$}
\newcommand{\ns}{--}
\newcommand{\cx}{$\times$}
\newcommand{\lwF}{\CIRCLE}     
\newcommand{\lwH}{\LEFTcircle} 
\newcommand{\lwE}{\Circle}     

\begin{document}

\pdfpagewidth=8.5in
\pdfpageheight=11in

\newcommand{\iscasubmissionnumber}{35}

\pagenumbering{arabic}

\title{Tureis: \textbf{T}ransformer-based \textbf{U}nified \textbf{Re}silience for \textbf{I}oT Devices in \textbf{S}mart Homes \vspace{-0.5em}}
\author{
\IEEEauthorblockN{
Alireza Borhani\IEEEauthorrefmark{1}
Vafa Andalibi\IEEEauthorrefmark{2},
Bahar Asgari\IEEEauthorrefmark{1}
}
\IEEEauthorblockA{\IEEEauthorrefmark{1}University of Maryland, College Park, MD, USA}
\IEEEauthorblockA{\IEEEauthorrefmark{2}Google, Manhattan, NY, USA}
\IEEEauthorblockA{\texttt{borhani@umd.edu}}
}
\IEEEaftertitletext{\vspace{-2.5em}}
\def\IEEEtitletopspaceextra{-0.6\baselineskip}
\maketitle
\thispagestyle{plain}
\pagestyle{plain}


\begin{abstract}
Smart-home IoT deployments rely on heterogeneous sensor networks whose correctness shapes application behavior and the physical environment. However, these sensors are low-cost and resource-constrained, and exposure to environmental stressors makes them highly prone to failure. Despite recent progress, existing methods remain limited: many assume single-failure, single-resident settings, offer only failure detection rather than sensor-level localization, or cover only a narrow range of fault types and sensor modalities. Others depend on labeled data and human intervention, or introduce substantial resource overhead that precludes deployment on resource-constrained edge devices. To overcome these limitations, we propose \textsc{Tureis}, a fully self-supervised, context-aware method for failure detection and faulty-sensor localization in smart homes, designed for multi-failure, multi-resident deployments on edge devices. \textsc{Tureis} encodes heterogeneous binary and numeric sensor streams into compact bit-level features. It then trains a lightweight BERT-style Transformer with sensor-wise masked reconstruction over short-horizon windows, capturing spatial and short-term temporal correlations without mixing in unrelated events. This self-supervised objective eliminates the need for labels or curated semantics. Then, at run-time, \textsc{Tureis} converts reconstruction residuals into sensor-level failure evidence and employs an iterative isolate-and-continue loop that masks flagged sensors. This iterative isolation allows additional failures to surface and enables resilient, fine-grained localization. Across five public smart-home datasets with up to nine residents, \textsc{Tureis} improves single-failure localization F1 score by \(+7.6\%\), \(+21.0\%\), and \(+25.0\%\) over three strong baselines. In multi-failure scenarios with up to five simultaneous faulty sensors, it further increases the localization F1 score by \(+17.6\%\) and \(+35.4\%\) over two baselines, while the third baseline does not extend to this setting. These gains come with short, minute-scale localization times and an edge-friendly footprint, as a sub-megabyte model that processes each minute of data in only a few milliseconds while keeping peak memory usage at roughly 0.5\,GB on a Raspberry~Pi~5.

\end{abstract}

\section{Introduction}
\label{intro}
\label{sec:intro}
The Internet of Things (IoT) has rapidly evolved from pilots to large-scale deployments in smart homes, cities, and healthcare systems, with 43 billion IoT devices expected by 2030~\cite{ericsson2025iot}. As these environments scale, heterogeneous sensor networks grow from a few to hundreds of devices per site~\cite{Catlett2017Arrayofthings, Teh2020Sensor}, turning the sensing layer into a high volume data source whose correctness directly governs application behavior.

IoT sensors are built from inexpensive electronic components, leading to limited durability and constrained resources that make them highly prone to failure~\cite{AlShehri2024AnEnergy-Efficient, Hnat2011Thehitchhikersguide}. In practice, wear out, battery issues, environmental stressors, and human factors commonly cause sensor failures~\cite{Moore2020IoT, Teh2020Sensor}. These failures have consequences that extend into the physical environment~\cite{Borhani2022ThingsDND} (e.g., a heater failing to turn off or a smoke detector suppressing alarms~\cite{Fu2021HAWatcher}) undermining the dependability of the entire system. Among IoT environments, smart homes exemplify this challenge, with dense deployments of heterogeneous, low cost sensors such as motion detectors, door contacts, water leakage detectors, and temperature sensors. The diversity and scale of these deployments have made preserving correct sensor operation a key concern, motivating extensive work on detecting faulty sensors in smart homes~\cite{Fu2021HAWatcher, Jung2022Simultaneous, Li2024SeIoT, DAngelo2023Transformer-based, Xiao2024Make, attarha2024assuresense, ElHady2020Sensor, Choi2018Detecting, Nguyen2021Effective, Mallick2020DETECTIF, Borhani2022ThingsDND}.

\vspace{-0.5pt}None of the existing methods, however, is comprehensive and suitable for deployment across the broad scope of real-world smart-home IoT systems, for several reasons. (i) Many methods assume a single-failure setting and struggle under simultaneous multi-failure cases. (ii) They are often limited to homes with one or two residents and degrade under multi-resident settings. (iii) Several methods have a high computational footprint, making deployment on resource-constrained edge devices impractical, as these devices have limited memory, power, and computational resources~\cite{Borhani2022Fast}. (iv) They cover only a narrow range of fault types or (v) a limited set of sensor modalities, which limits their generality in heterogeneous homes. (vi) Methods often are detection-only and lack sensor-level localization, while localization is necessary to isolate and accommodate faulty sensors and to maintain resilient downstream operation by preventing cascading errors and incorrect actions~\cite{Xing2021Cascading}. Lastly, (vii) many methods require labels or human-provided semantics, which contradicts the IoT goal of eliminating human intermediaries~\cite{vermesan2019next, Fizza2023ASurvey}. Together, no prior work simultaneously meets all requirements (i)–(vii), which makes them inappropriate to use in smart-home environments.

\vspace{-0.5pt}To address these limitations, we propose \textsc{Tureis}, a transformer-based context-aware method, for failure detection and faulty sensor localization in smart homes that meets all requirements (i)–(vii). Our central hypothesis is that a short-horizon, sensor-wise masked BERT-style Transformer is structurally well-suited to this problem for four reasons. \textbf{\textit{First}}, smart-home sensors exhibit structured spatial (i.e., cross-sensor) and short-term temporal correlations, and many fault types do not yield large univariate deviations but instead violate these correlations. Self-attention inherently aggregates interactions across all sensors and time intervals within the context, and bidirectional self-attention leverages both preceding and subsequent intervals, effectively capturing spatiotemporal correlations of sensors. \textbf{\textit{Second}}, multi-head self-attention factorizes interactions into a small number of correlation subspaces, and thus, in the multi-failure scenario, disruption in one subspace can be contrasted with stability in others, improving separability and yielding clean failure signals. Moreover, multiple layers compose the subspaces through depth, helping learn subtle spatiotemporal correlations that are easily diluted in multi-resident homes. \textbf{\textit{Third}}, training with sensor-wise masked reconstruction makes localization the learning objective, and the transformer must reconstruct a masked sensor from its neighboring sensors and temporal context, requiring no labels or curated semantics. Then, the resulting per-sensor reconstruction loss serves as failure evidence, enabling faulty sensor localization rather than failure detection only. \textbf{\textit{Fourth}}, training under sensor-wise masking enables an isolate-and-continue loop: once a sensor is located, it is masked going forward, keeping inputs in-distribution, suppressing interference from already-faulty sensors, and allowing additional failures to surface sequentially.

\textsc{Tureis} first constructs compact, bit-level, early-fused features from raw, heterogeneous sensor streams, unifying diverse sensor channels into a common representation. These bit-level encodings reduce on-device data movement compared to floating-point representations, which are dominant contributors to energy use in IoT systems~\cite{Jahannia2024AnEnergy, Jahannia2025DaLAMED}. Then, \textsc{Tureis} trains a lightweight encoder-only BERT-style Transformer with sensor-wise masked reconstruction. Together, the bit-level early fusion and the lightweight Transformer encoder support multiple sensor types while keeping the resource footprint low. Finally, at run-time, it converts reconstruction residuals into per-sensor evidence and uses an isolate-and-continue loop to surface simultaneous failures.

We evaluate \textsc{Tureis} on five public smart-home datasets with up to nine residents \cite{Cook2013CASAS, vankasteren2011activity}, and compare it against three baselines, ThingsDND \cite{Borhani2022ThingsDND}, DICE \cite{Choi2018Detecting} and Anomaly Transformer (AT) \cite{Xu2022Anomaly}. The results show that when there is one faulty sensor, \textsc{Tureis} raises the average F1 score for faulty sensor localization by \(+7.6\%\), \(+25.0\%\), and \(+21.0\%\) over ThingsDND, DICE, and AT, respectively. For multiple simultaneous failures, it improves the localization F1 score by \(+17.6\%\) and \(+35.4\%\) over ThingsDND and DICE, respectively (AT does not support multi-failure localization). Moreover, in terms of responsiveness, \textsc{Tureis} locates the faulty sensor \(10.81\times\) and \(10.91\times\) faster than ThingsDND and DICE, respectively, and is \(2.93\times\) slower than AT while still maintaining minute-scale localization time. Additionally, these gains come with minimal overhead and a footprint suitable for edge devices. The compact encoder has an average of \(0.05\) million parameters and operates in real-time on a Raspberry Pi 5 board, taking only \(1.46\)-\(4.12\) milliseconds to process one minute of data, with peak run-time memory usage of roughly \(0.5\) GB.

\section{Background and Motivation}

\label{BackRW}
\label{sec:BackRW}

\subsection{Sensor Faults}
\vspace{-3pt}
Sensor data faults fall into two categories \cite{Choi2018Detecting}: fail-stop and non-fail-stop faults. Fail-stop faults lead to the shutdown of the sensor, while non-fail-stop faults only generate incorrect data. The most commonly observed non-fail-stop faults are outlier, spike, stuck-at, high-noise, and drift~\cite{Ni2009Sensor}.

\textbf{\textit{Outlier:}} 
It is an isolated sensor reading that temporarily or spatially deviates from the expected range. Let $k$ denote the time index (sample step) and $x[k]$ the original reading. Formally, the outlier-affected sensor readings are:
\vspace{-5pt}
\begin{equation}
\tilde{x}_{\text{outlier}}[k] =
\begin{cases}
o, & k = \kappa, \; o \notin I[\kappa] \\
x[k], & \text{otherwise}
\end{cases}
\end{equation}
where $\kappa$ is the index at which the outlier occurs, $o$ is the outlier value, 
and $I[\kappa]$ denotes the expected range of valid readings at step $\kappa$, 
typically estimated from a temporal window of past samples or from spatially correlated sensors.

\textbf{\textit{Spike:}} 
It is a rate of change much greater than expected over a short time. A spike consists of consecutive readings that deviate sharply due to unusually large gradients. Formally, the spike-affected sensor readings are represented as: 
\vspace{-5pt}
\begin{equation}
\tilde{x}_{\text{spike}}[k] =
\begin{cases}
x[k] + \delta[k], &
  \substack{\tau \le k < \tau + \Delta,} \\
x[k], & \text{otherwise}
\end{cases}
\end{equation}
where $\tau$ is the index at which the spike begins, $\Delta$ is the duration of the spike, 
$\delta[k]$ is the sudden change applied at each step during the spike, which drives the discrete gradient far beyond the expected range estimated from temporal windows or from spatially correlated sensors.

\textbf{\textit{Stuck-at:}} 
It is a series of sensor readings that remain constant for a while longer than expected, resulting in experiencing zero variation during that period. Formally, the stuck-at-affected sensor readings are represented as:
\vspace{-5pt}
\begin{equation}
\tilde{x}_{\text{stuck-at}}[k] =
\begin{cases}
s, & \tau \leq k < \tau + \Delta \\
x[k], & \text{otherwise}
\end{cases}
\end{equation}
where $\tau$ is the index at which the fault begins, $\Delta$ is the duration of the fault, and 
$s$ is a fixed sensor value which can lie within or outside the expected operating range.

\textbf{\textit{High-noise:}} 
It is an unusually high amount of noise or variance in sensor values. Formally, the high-noise-affected sensor readings are represented as:
\vspace{-5pt}
\begin{equation}
\tilde{x}_{\text{noise}}[k] =
\begin{cases}
x[k] + \epsilon[k], & \substack{\tau \leq k < \tau + \Delta,} \\
x[k], & \text{otherwise}
\end{cases}
\end{equation}
where $\tau$ is the index at which the fault begins, $\Delta$ is the duration of the noisy interval,  $\epsilon[k]$ is an additive fluctuation at step $k$ that inflates the windowed variance beyond the expected range estimated from historical or neighboring readings.

\textbf{\textit{Drift:}} 
A drift is a gradual deviation in sensor readings that develops slowly and progressively over time. Formally, the drift-affected sensor readings are represented as:
\vspace{-5pt}
\begin{equation}
\tilde{x}_{\text{drift}}[k] =
\begin{cases}
x[k] + d(k-\tau), & \tau \leq k < \tau + \Delta \\
x[k], & \text{otherwise}
\end{cases}
\end{equation}
where $\tau$ is the index at which the drift begins, $\Delta$ is the duration of the drift, 
and $d(k-\tau)$ is a monotonic function that introduces a progressive shift over time.

\begin{table*}[htbp]
\centering
\caption{Comparison of Prior Work on Sensor Failure Detection \& Localization in Smart Homes and Related Domains.}
\vspace{-5pt}
\label{tab:comparison_updated}
\resizebox{\textwidth}{!}{
\small
\begin{tabular}{l | c | c | c c | c c c c c c | c c c | l}
\toprule
\textbf{Work} & \vhead{Domain} & \vhead{Lightweight} & \vhead{Multi-failure} & \vhead{Multi-resident} & \vhead{Fail-stop} & \vhead{Outlier} & \vhead{Spike} & \vhead{Stuck-at} & \vhead{High-noise} & \vhead{Drift} & \vhead{Localization} & \vhead{Heterogeneous} & \vhead{Unlabeled} & \textbf{Method} \\
\midrule
Mallick et al.~\cite{Mallick2020DETECTIF} & SH & \lwH & \cx & \cx & \ck & \ck & \ck & \ck & \cx & \cx & \ck & \ck & \cx & SEM (Freq. Itemsets + Spatiotemp. Rules) \\
ElHady et al.~\cite{ElHady2020Sensor} & SH & \lwF & \cx & \cx & \ck & \cx & \cx & \cx & \cx & \cx & \cx & \cx & \ck & SEM (Assoc. Rule Mining) \\
Fu et al.~\cite{Fu2021HAWatcher} & SH & \lwE & \cx & \cx & \cx & \ck & \cx & \cx & \cx & \cx & \cx & \ck & \cx & SEM (Semantic Corr. + Shadow Exec.) \\
Nguyen et al.~\cite{Nguyen2021Effective} & SH & \lwF & \cx & \cx & \cx & \ck & \cx & \cx & \cx & \cx & \cx & \cx & \cx & SEM (Spatial/Depend. Corr.) \\
Li et al.~\cite{Li2024SeIoT} & SH & \lwF & \ck & \cx & \ck & \ck & \ck & \ck & \cx & \cx & \ck & \ck & \cx & SEM (GNN + Attn.) \\
Xiao et al.~\cite{Xiao2024Make}  & SH & \lwH & \cx & \cx & \cx & \cx & \cx & \ck & \ck & \cx & \cx & \ck & \ck & SEM (Transformer + Loss-Guided Mask) \\
Choi et al.~\cite{Choi2018Detecting} -- \textbf{DICE*} & SH & \lwF & \cx & \cx & \ck & \ck & \ck & \ck & \ck & \cx & \ck & \ck & \ck & CA (Sensor Corr. + Markov Chks.) \\
Jung et al.~\cite{Jung2022Simultaneous} & SH & \lwH & \ck & \cx & \cx & \ck & \cx & \cx & \cx & \cx & \ck & \cx & \ck & CA (DNN + Hypersphere) \\
Borhani et al.~\cite{Borhani2022ThingsDND} -- \textbf{ThingsDND*} & SH & \lwF & \cx & \ck & \ck & \ck & \ck & \ck & \ck & \cx & \ck & \ck & \ck & CA (BiLSTM) \\
Attarha et al.~\cite{attarha2024assuresense} & SH & \lwF & \cx & \cx & \ns & \ns & \ns & \ns & \ns & \ns & \cx & \ck & \cx & CA (Spatiotemp. Corr. + Classifier) \\
Xu et al.~\cite{Xu2022Anomaly} -- \textbf{AT*} & IIoT & \lwE & \cx & \ck & \ck & \ck & \ck & \ck & \ck & \ck & \ck & \cx & \ck & XSeq (Transformer + Assoc. Discrepancy) \\
D'Angelo et al.~\cite{DAngelo2023Transformer-based} & SH & \lwH & \cx & \ck & \ck & \cx & \cx & \ck & \cx & \cx & \cx & \cx & \ck & XSeq (Transformer + RoBERTa) \\
Darvishi et al.~\cite{Darvishi2023Deep} & IIoT & \lwE & \ck & \ns & \cx & \cx & \cx & \cx & \ck & \ck & \ck & \ck & \cx & XSeq (GCRN + MLP) \\
Kayan et al.~\cite{Kayan2024CASPER} & CPS & \lwF & \cx & \cx & \ck & \cx & \cx & \cx & \cx & \ck & \cx & \cx & \cx & XSeq (1D-CNN) \\
Dong et al.~\cite{Dong2024Real-time} & IIoT & \lwF & \cx & \cx & \ns & \ns & \ns & \ns & \ns & \ns & \cx & \ck & \cx & XSeq (Attn-LSTM, GA-Tuned) \\
\midrule
\textbf{Our Work (\textsc{Tureis})} & \textbf{SH} & \textbf{\lwF} & \textbf{\ck} & \textbf{\ck} & \textbf{\ck} & \textbf{\ck} & \textbf{\ck} & \textbf{\ck} & \textbf{\ck} & \textbf{\ck} & \textbf{\ck} & \textbf{\ck} & \textbf{\ck} & \textbf{CA (Self-Sup. BERT-Style Transformer)} \\
\bottomrule
\addlinespace[2pt]
\bottomrule
\addlinespace[2pt]
\multicolumn{15}{l}{\footnotesize \ck: Yes, \cx: No, \ns: Not specified $|$ \lwF: Low, \lwH: Moderate, \lwE: High $|$ \textbf{SH}: Smart Home, \textbf{IIoT}: Industrial IoT, \textbf{CPS}: Cyber-Physical System.} \\
\multicolumn{15}{l}{\footnotesize \textbf{CA}: Context-Aware, \textbf{SEM}: Semantics-Based, \textbf{XSeq}: Cross-Domain Sequence Model $|$ \textbf{* DICE, ThingsDND,} and \textbf{AT}: our baselines for comparison.} \\
\end{tabular}
}
\vspace{-12pt}
\end{table*}

\subsection{Prior Sensor-Failure Detection Methods}
\vspace{-3pt}
Prior work on sensor failure and anomaly detection in smart homes and related IoT domains falls into three main categories: semantics based methods~\cite{Mallick2020DETECTIF, Fu2021HAWatcher, ElHady2020Sensor, Xiao2024Make, Nguyen2021Effective, Li2024SeIoT}, context-aware methods~\cite{Choi2018Detecting, Borhani2022ThingsDND, Jung2022Simultaneous, attarha2024assuresense}, and cross-domain sequence models~\cite{DAngelo2023Transformer-based, Darvishi2023Deep, Dong2024Real-time, Kayan2024CASPER, Xu2022Anomaly}. We compare representative studies in Table~\ref{tab:comparison_updated} and review each in turn, focusing on their design and limitations relative to our objectives.

\subsubsection*{\textbf{Semantics-based methods}}
These methods explicitly encode application or environment semantics. Mallick et al.~\cite{Mallick2020DETECTIF} addresses transient faults focusing on activities of daily living (ADLs) by mining frequent item sets and spatiotemporal rules. It distinguishes missing versus spurious events and can repair noisy streams. However, it relies on ADL labels, is evaluated on single-resident homes under single-fault scenarios, and does not optimize for resource-constrained gateways. ElHady et al.~\cite{ElHady2020Sensor} propose a failure detection system for Ambient Assisted Living using association rule mining, deriving strong sensor-sensor correlations from unlabeled data, then maintaining per-sensor health scores and flagging failures when scores drop below a threshold. While it effectively uses unlabeled data, it is limited to binary sensors in a single-resident home and is evaluated under single-failure settings.

Fu et al.~\cite{Fu2021HAWatcher} and Nguyen et al.~\cite{Nguyen2021Effective} leverage application and deployment semantics for anomaly detection. Fu et al. builds explainable correlations from app code, device types, and installation locations and performs semantics-aware shadow execution, while Nguyen et al. combine spatial and dependable correlations with resident-count–based reasoning to flag anomalies. However, both are evaluated on single anomaly in single-resident homes, with Fu et al. incurring notable overhead from shadow execution and Nguyen et al. focusing on binary sensors. SeIoT~\cite{Li2024SeIoT} and SmartGuard~\cite{Xiao2024Make} use knowledge graphs and behavior modeling, respectively, to detect anomalies. However, both are limited to single-resident settings, with SeIoT relying on labeled network-level attack traces and SmartGuard targeting only single anomalies.
\subsubsection*{\textbf{Context-aware methods}}
These methods infer spatial and/or temporal correlations directly from sensor data and their surrounding context. DICE~\cite{Choi2018Detecting} learns sensor correlations and transition probabilities offline and performs run-time correlation and transition checks for faulty sensor detection and localization. It supports both binary and numeric sensors, requires no labels, has low computational overhead, and covers fail-stop and several non–fail-stop fault types. However, it is designed for single-failure, single-resident setting. Jung et al.~\cite{Jung2022Simultaneous} employ a hypersphere-based DNN to learn robust decision boundaries around normal data, requiring a small set of anomalies for calibration and maintaining performance under simultaneous anomalies. However, it focuses only on binary sensors in a single-resident home and targets only outlier-style anomalies.

Borhani et al.~\cite{Borhani2022ThingsDND} (ThingsDND) extends this direction with a lightweight BiLSTM-based framework that learns spatiotemporal correlations from unlabeled data, predicts sequences, and increments per-sensor failure scores until thresholds are crossed. 
ThingsDND comes very close to our target setting, but it still assumes one faulty sensor at a time and does not cover drift faults. Attarha et al.~\cite{attarha2024assuresense} develops a feature extractor that captures spatiotemporal sensor correlations and then feeds them to a lightweight classifier. It is designed for low-power edge devices, but it relies on measurements and cannot accommodate multi-resident, multi-failure scenarios. Moreover, it does not distinguish specific data-fault types and is evaluated based on faulty measurements arising from sensing material degradation and external interference. In short, context-aware methods are close to our objectives as they operate directly on raw sensor streams, exploit spatiotemporal correlations without requiring rich semantics, and can be made lightweight and unlabeled. 
\subsubsection*{\textbf{Cross-domain sequence models}}
A third line of work adapts generic time-series architectures often developed outside the smart homes domain (e.g., industrial IoT and cyber-physical systems). Anomaly Transformer~\cite{Xu2022Anomaly} (AT) is an unsupervised model that learns prior and series associations via a specialized anomaly-attention mechanism. It separates normal and anomalous data by maximizing an association discrepancy metric, computed as the KL divergence between the learned series-association and a fixed Gaussian-based prior association. AT achieves state-of-the-art results on several continuous multivariate time-series benchmarks. However, it is designed for continuous-valued time series data, rather than for event-driven sensors commonly found in smart homes. Moreover, it employs a complex architecture that is not optimized for resource-constrained gateways and does not address multi-failure scenarios. D’Angelo et al.~\cite{DAngelo2023Transformer-based} introduces a Transformer-based Compound Correlation Miner that learns two-to-one event correlations (e.g., for stealthy attacks) using a RoBERTa model trained on normal event logs. However, it does not detect multiple simultaneous failures in event-driven sensor values. 

Darvishi et al.~\cite{Darvishi2023Deep} proposes a deep recurrent GCN with a denoising training scheme and a classification block to perform fault detection, isolation, and accommodation in digital twins with millisecond-scale latency. It relies on labeled data, and its computational complexity grows exponentially with the size of the system network. Kayan et al.~\cite{Kayan2024CASPER} applies 1D-CNN models on edge boards for robotic arm motion anomaly detection, and Dong et al.~\cite{Dong2024Real-time} design GA–Att–LSTM, a genetic algorithm–optimized attention-LSTM for industrial IoT fault detection. Both methods demonstrate edge feasibility, but Kayan et al. is tailored to continuous signals rather than event-driven sensors. Both approaches rely on labeled anomalies or domain-specific supervision and do not address scenarios involving simultaneous multi-failures or multi-anomalies.

In summary, semantics-based, context-aware, and cross-domain sequence models have advanced reliable sensing, but a critical gap remains. No prior method simultaneously handles multiple sensor failures in multi-resident homes, covers diverse data fault types on heterogeneous binary and numeric sensors, operates fully unlabeled, and stays lightweight enough for continuous deployment on resource constrained home gateways. These limitations motivate the design of \textsc{Tureis}.


\section{Tureis: The Proposed Method}
\label{Tureis}
\label{sec:tureis}

\subsection{Key Insights}
\vspace{-3pt}
To address the challenge and meeting our design goals, \textsc{Tureis} introduces an automatic, context-aware method that focuses on bit-level feature encoding and a self-supervised learning signal, which together make unlabeled detection tractable. To do this, the key insight of \textsc{Tureis} is to employ a short spatiotemporal context to capture activity chains without drifting into unrelated events, since each user activity in a smart home triggers a specific set of sensors. Since there are no labels available, the model is trained through masked reconstruction. Specifically, for each sequence, it masks out an entire sensor's data throughout the sequence and a BERT-style Transformer encoder learns to reconstruct it from the remaining sensors and the temporal context. This approach forces the model to internalize both cross-sensor and temporal structure that characterize normal operation. 

Moreover, bidirectional self-attention with positional embeddings enables the encoder to attend to past and future within the sequence simultaneously. This capability allows \textsc{Tureis} to integrate subtle, distributed correlations in multi-resident homes. Furthermore, bit-level features keep the input small and maintain memory and bandwidth low, and a lightweight encoder-only Transformer with two layers and four attention heads makes the method practical for edge deployment. Finally, in inference, failures surface as deviations in reconstruction, and to handle simultaneous failures, \textsc{Tureis} runs an iterative isolation loop. Specifically, when a sensor trips, it suppresses the sensor's feature stream from that time onward, eliminating interference and enabling the localization of further failures. 
\vspace{-3pt}
\subsection{Pipeline Overview}
\begin{figure}[t]
    \centering
    \includegraphics[width=1\linewidth,trim=175 550 125 160,clip]{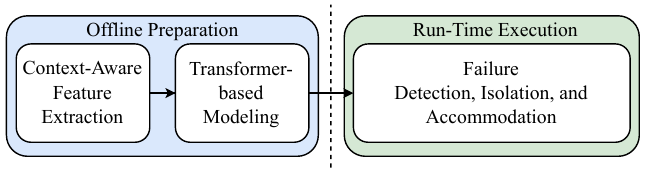}
    \vspace{-20pt}
    \caption{\textbf{\textsc{Tureis} overview:} An offline phase performs feature extraction and trains a BERT-style encoder via sensor-wise masked reconstruction. Then, a run-time execution phase applies the model to each window to derive per-sensor failure scores and iteratively mask tripped sensors for resilient operation.}
    \vspace{-17pt}
    \label{fig:Tureis-Overview}
\end{figure}

\vspace{-3pt}
As illustrated in Figure~\ref{fig:Tureis-Overview}, \textsc{Tureis} has two phases: 1) an \textit{offline} preparation phase that includes a context-aware feature extraction module and a transformer-based modeling module, and 2) a \textit{run-time} execution phase that comprises a streaming inference module for resilient operation. In the feature extraction module, heterogeneous sensor streams are aggregated into uniform intervals, each channel is summarized into compact features, and all channels are fused at the feature level (i.e., early fusion) to create a single windowed input for the model. In the model training module, a lightweight BERT-style encoder learns spatio-temporal correlations from unlabeled data through sensor-wise masked reconstruction. Then, the streaming inference module employs the trained model on each new window, converts reconstruction residuals into per-sensor failure scores, and when a sensor trips, it masks that channel in the following windows to suppress its impact, surface additional failures, and accommodate using the remaining healthy sensors.
\vspace{-12pt}
\subsection{Offline Preparation: Feature Extraction \& Model Training}

\subsubsection*{\textbf{Context-Aware Feature Extraction}}
In this module, heterogeneous sensor streams are aggregated into fixed-length time intervals that keep the context from multiple sensors while staying compact. First, for each interval, features from every channel are summarized into compact features. Then, these features from each channel are concatenated in an early fusion process to form a single interval feature vector. Specifically, let \(S_{k,\tau}\) denote the set of values produced by sensor \(k\) within interval \(\tau\): \(S_{k,\tau} = \{\, s_{k,\tau,i} \mid i = 1,\dots, m_{k,\tau} \,\}\), where \(m_{k,\tau} = \lvert S_{k,\tau}\rvert\) is the number of emissions from sensor \(k\) during that time interval, \(\tau\) indexes the interval, \(k\) indexes the sensor channel, and \(s_{k,\tau,i}\) is the \(i\)-th observed emission of sensor \(k\) within the interval. Then, for each interval \(\tau\), data from each channel is summarized into just a few bits.

For binary channels, we encode two activity bits to capture per-interval activity intensity. To achieve this, per-sensor thresholds \(P_{25,k}\) and \(P_{75,k}\) are calculated, which represent the 25th and 75th percentiles of per-interval activation counts during training. This helps categorize activity into low, medium, and high. Specifically, the two bits \((b^{(1)}_{k,\tau}, b^{(2)}_{k,\tau})\) are encoded as:
\vspace{-5pt}
\begin{equation}
(b^{(1)}_{k,\tau}, b^{(2)}_{k,\tau}) =
\begin{cases}
(0,0), & m = 0,\\
(0,1), & 0 < m < P_{25,k},\\
(1,0), & P_{25,k} \le m < P_{75,k},\\
(1,1), & m \ge P_{75,k}.
\end{cases}
\end{equation}

For numeric channels, we emit four bits per interval. These include the same two activity bits (i.e., \(b^{(1)}_{k,\tau}, b^{(2)}_{k,\tau}\)) following the same rule as for binary channels, along with two dynamics bits that capture short-term variability within the interval. Let the consecutive differences be \(\Delta s_{k,\tau,\ell} = s_{k,\tau,\ell} - s_{k,\tau,\ell-1}\) where \(\ell=2,\dots,m\). Then we define these two dynamic bits as:
\vspace{-5pt}
\begin{equation}
\label{eq:jumpy}
j_{k,\tau} =
\begin{cases}
1, & \operatorname{std}\!\big(\Delta s_{k,\tau,2{:}m}\big) > \sigma_k^{\Delta},\\[4pt]
0, & \text{otherwise}
\end{cases}
\end{equation}

\begin{equation}
\label{eq:burst}
u_{k,\tau} =
\begin{cases}
1, & \max\!\big(\Delta s_{k,\tau,2{:}m}\big) > \operatorname{med}_k^{\Delta},\\[2pt]
0, & \text{otherwise}
\end{cases}
\end{equation}
where \(\sigma_k^{\Delta}\) is the per-sensor baseline volatility, computed during training as the standard deviation of all successive differences \((\Delta s_{k,\tau,\ell})\) for sensor \(k\). The term \(\operatorname{med}_k^{\Delta}\) is the per-sensor typical step size, determined as the median of the absolute successive differences \(\bigl|\Delta s_{k,\tau,\ell}\bigr|\) for sensor \(k\) over the training set. Intuitively, the Jumpy bit in \eqref{eq:jumpy} activates when the volatility within the interval exceeds the baseline noise of sensor, and the Burst bit in \eqref{eq:burst} activates when the largest single step within the interval exceeds the typical step size of sensor; otherwise, both bits are \(0\).

Therefore, a numeric channel gives a four-bit encoding (i.e., \(b^{(1)}_{k,\tau}, b^{(2)}_{k,\tau}, j_{k,\tau}, u_{k,\tau}\)), while a binary channel provides two bits (i.e., \(b^{(1)}_{k,\tau}, b^{(2)}_{k,\tau}\)). Then, concatenating all per-channel encoding yields the interval feature vector \(x_\tau \in \{0,1\}^{D}\), where \(D\) is the total number of bits after early fusion (i.e., two for each binary sensor and four for each numeric sensor). Consequently, adjacent interval vectors are then organized into short windows of length \(L\) with a unit stride:
\[
X_{\tau:\,\tau+L-1}
= \big[\,x_{\tau},\,x_{\tau+1},\,\dots,\,x_{\tau+L-1}\,\big]
\in \{0,1\}^{L\times D}.
\]
This early-fused, windowed multivariate representation exposes cross-sensor and context to the model while keeping the footprint small.

\begin{figure}[t]
  \centering
  \includegraphics[width=\linewidth]{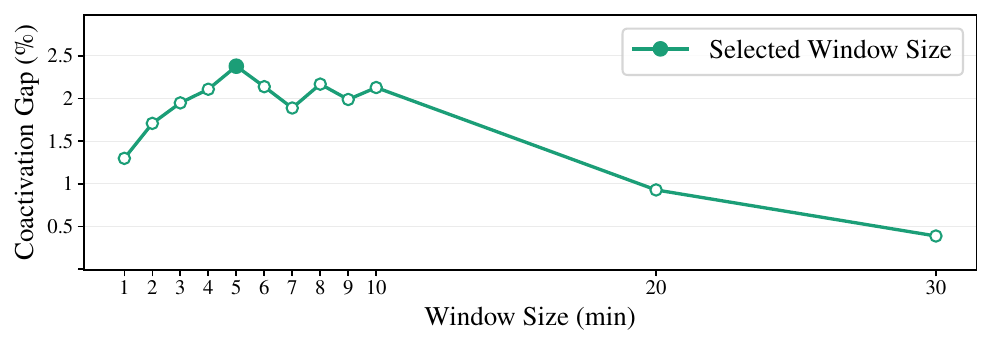}
  \vspace{-23pt}
  \caption{\textbf{Coactivation gap between correlated and uncorrelated sensor pairs versus window length $L$:} The curve peaks at five minutes, and thus, $L=5$ is selected for all modules.}
  \vspace{-15pt}
  \label{fig:mi-gap}
\end{figure}

To select an optimal window length $L$, we measure how well a time window captures sensor correlations. Specifically, for each $L$, we compute a \emph{Coactivation Gap}, defined as the difference in the coactivation probability between correlated sensor pairs (e.g., in the same room or historically active together) and uncorrelated pairs. Here, coactivation means that both sensors are active at least once within the window. The gap increases for short windows, peaks at five minutes, and then decreases as longer windows mix unrelated events (Figure~\ref{fig:mi-gap}). Therefore, we set $L=5$ for all modules.

Moreover, we set the interval $\tau$ length to one minute and use a stride of one interval. With this choice, consecutive windows have high overlap (i.e., 80\%), which means that delayed but related coactivations, up to four minutes apart, still appear in at least one window. If, instead, the interval were shorter than one minute, each window ($L=5$) would contain more interval feature vectors, which would increase the attention computation cost. Additionally, it would create overlapping, nearly duplicate contexts that encourage memorization rather than generalization. In contrast, larger strides, such as 2-3 minutes, spread these events across different windows, increasing the chances of missing detections or delays. Therefore, we choose a one-minute interval length as a canonical integer-aligned option that balances computation, stability, and timely detection.

\begin{figure}[t]
    \centering
    \includegraphics[width=1\linewidth,trim=40 670 260 50,clip]{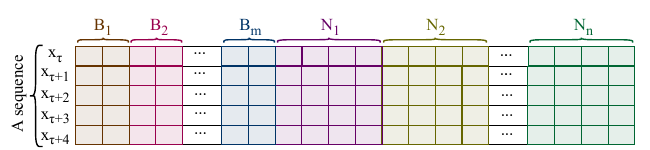}
    \vspace{-20pt}
    \caption{\textbf{Five-interval window (i.e., sequence) produced by the context-aware feature extraction module:} Rows are consecutive interval feature vectors $\in\{0,1\}^{D}$ obtained via early fusion across channels, $B_1,\ldots,B_m$ denote binary sensors (2 bits per sensor) and $N_1,\ldots,N_n$ denote numeric sensors (4 bits per sensor).}
    \vspace{-20pt}
    \label{fig:A-sequence}
\end{figure}

With a one-minute interval and a window length of $L=5$, Figure~\ref{fig:A-sequence} shows a single window (i.e., a sequence). In the figure, columns labeled \texttt{B} represent binary sensors, columns labeled \texttt{N} represent numeric sensors, and the rows correspond to the five consecutive interval feature vectors. This window, and each subsequent window in the stream, serves as an input to the next module. Specifically, the same window forms a length-$L$ sequence for the encoder, where its interval vectors are projected to embeddings, and a lightweight BERT-style encoder is trained through sensor-wise masked reconstruction.

\subsubsection*{\textbf{Transformer-based Modeling}}
\begin{figure*}[t]
    \centering
    \includegraphics[width=0.8\textwidth, trim=130 540 60 80,clip]{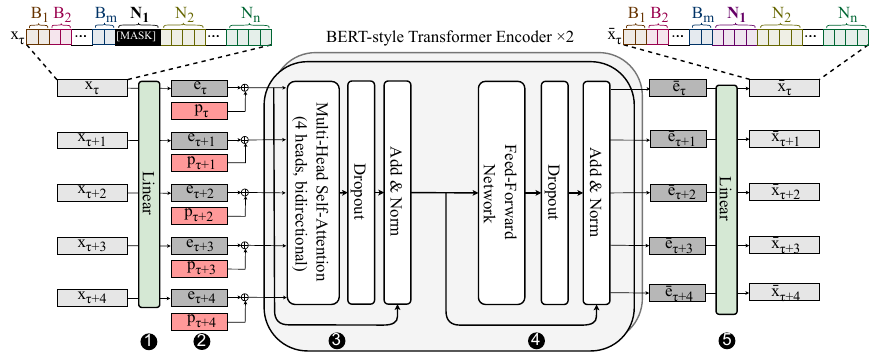}
    \vspace{-10pt}
    \caption{\textbf{Transformer-based Modeling:} masked reconstruction in \textsc{Tureis}. In this illustration, the numeric channel $N_1$ is masked across all $L$ intervals. However, in general, each training sequence masks a random subset of sensor channels by replacing their bit positions with a learnable \texttt{[MASK]} token.}
    \vspace{-15pt}
    \label{fig:transformer}
\end{figure*}

As illustrated in Figure~\ref{fig:transformer}, given a window $X_{\tau-L+1:\,\tau}\in\{0,1\}^{L\times D}$, which is treated as a sequence for the encoder, this module learns spatiotemporal correlations of normal operation from unlabeled data through sensor-wise masked reconstruction. In other words, the model is trained to reconstruct masked sensor channels using the remaining context in the window.

Within each window, each interval vector $x_{\tau}\in\{0,1\}^{D}$ is mapped to an embedding $e_{\tau}\in\mathbb{R}^{d}$ using a learnable linear projection $e_{\tau}=W_{\text{in}}x_{\tau}+b_{\text{in}}$ (\ncirc{1}). This embedding combines with a learnable absolute positional vector $p_{\tau}\in\mathbb{R}^{d}$ to form the position-aware token (\ncirc{2}). Here, $d$ is a compact embedding width chosen to balance efficiency and head partitioning. We set the per-head width to 16, which is a practical lower bound for stable, expressive attention. With four heads, we obtain $d=4\times16=64$. This choice helps keep parameter counts and computations low, avoiding both the under-capacity issues of tiny heads and the higher cost of larger $d$, which offer limited benefits for short windows.

We utilize a compact, BERT-style encoder-only Transformer with two encoder layers and four attention heads per layer. Specifically, each layer applies multi-head self-attention, dropout, residual connection, and layer normalization (\ncirc{3}). It is followed by a position-wise feed-forward network with widths $d\!\rightarrow\!2d\!\rightarrow\!d$ and a GELU activation, then dropout, residual connection, and layer normalization (\ncirc{4}). The encoder produces a contextual embedding $\hat e_{\tau}\in\mathbb{R}^{d}$ for each interval $\tau$ in the window. A learnable linear head projects each token back to the original bit space: $\hat x_{\tau}=W_{\text{out}}\hat e_{\tau}+b_{\text{out}}\in\mathbb{R}^{D}$(\ncirc{5}).

For each training sequence, we replace a random set of sensor channels with a learned \texttt{[MASK]} token across all $L$ intervals. The encoder receives this partially masked sequence and is trained to reconstruct the masked channels using the remaining sensors and the temporal context within the window. The training loss applies only to the masked positions and employs focal loss ~\cite{Lin2017Focal}, and optimization uses Adam. This approach reduces the impact of abundant inactivity and highlights short activations, preventing the signal from being dominated by the many inactive bits. Therefore, unmasked elements serve solely as context and incur no loss, and this channel-wise masking forces the model to understand cross-sensor correlations and temporal dependencies while remaining completely label-free. Then, at run-time, reconstruction residuals are converted into per-sensor failure scores, and an iterative isolation procedure handles multi-failure cases.

\subsection{Run-Time Execution: Streaming Inference for Resilience}
\vspace{-3pt}
At run-time, \textsc{Tureis} uses the same interval length, stride, and early-fusion layout as it did during the offline phase. It processes a rolling stream of feature vectors, and as shown in Figure~\ref{fig:postprocessing}, at each interval $\tau$, it forms a length-$L$ sequence from the most recent $L$ intervals: $X_{\tau:\,\tau+L-1}\in\{0,1\}^{L\times D}$ (\pcirc{1}). Then, the trained encoder maps this sequence to logits $\hat X_{\tau:\,\tau+L-1} \in \mathbb{R}^{L\times D}$ (\pcirc{2}), which are converted to per-sensor failure evidence as follows. For sensor $k$, let $I_k$ denote the bit indices that correspond to that sensor in the fused interval vector. A scalar reconstruction residual for $k$ at sequence $\tau$ is obtained by averaging the element-wise logistic reconstruction loss over all $L$ intervals and over the bits of that sensor:

\begin{figure}[t]
    \centering
    \includegraphics[width=\linewidth,trim=80 570 175 50,clip]{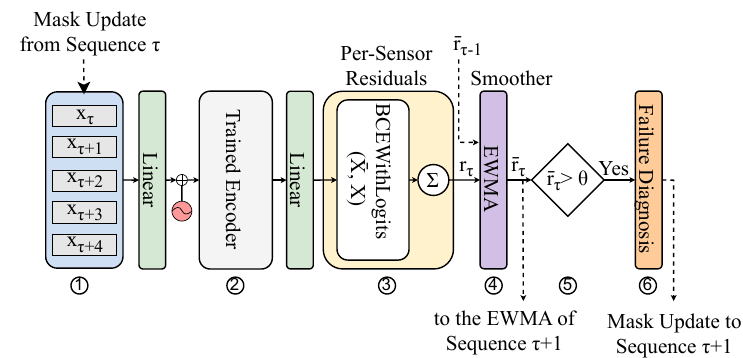}
    \vspace{-15pt}
    \caption{\textbf{Run-time execution and streaming inference in \textsc{Tureis}:} Reconstruction logits are converted to per-sensor residuals, EWMA-smoothed, and compared to calibrated per-sensor thresholds for failure localization. Flagged sensors are then masked in subsequent sequences to prevent interference and maintain uninterrupted streaming.}
    \vspace{-17pt}
    \label{fig:postprocessing}
\end{figure}

\vspace{-5pt}
\begin{equation}
\label{eq:residual-runtime}
r_{k,\tau}
= \frac{1}{L\,|I_k|}
\sum_{t=\tau}^{\tau+L-1}\ \sum_{j\in I_k}
\mathrm{BCE}\!\bigl(\hat X_{t,j},\,X_{t,j}\bigr)
\end{equation}
where $|I_k|$ can be either 2 or 4 since, during feature extraction, binary sensors are encoded with two bits and numeric sensors with four bits. $\mathrm{BCE}$ denotes the standard binary cross-entropy between the predicted logit and the actual observed bit, and $r_{k,\tau}$ represents the residual per sensor per window (\pcirc{3}).

These residuals serve as evidence of failure for each sensor. However, since these instantaneous residuals may fluctuate, we smooth them using an exponentially weighted moving average (EWMA) to improve stability (\pcirc{4}):
\vspace{-5pt}
\begin{equation}
\label{eq:ewma}
\hat{r}_{k,\tau} = \alpha\, r_{k,\tau} + (1-\alpha)\, \hat{r}_{k,\tau-1},
\qquad \alpha \in [0,1].
\end{equation}
Here $\alpha$ is the smoothing factor that controls the memory of EWMA. A larger $\alpha$ puts more focus on the current residual, while a smaller $\alpha$ allows longer memory. $\alpha$ is determined by a half-life, the time it takes for a residual’s weight to reduce to halfand is set to the sequence length $L$ indicated by the Coactivation Gap (Figure~\ref{fig:mi-gap}). By matching the half-life with $L$, we connect the smoother’s memory to the model’s correlation horizon, which dampens brief spikes and lets longer-lasting deviations within the $L$-minute window increase quickly.

Once the smoothed residuals $\hat{r}_{k,\tau}$ are obtained, we need a baseline for comparison, which is computed before run-time using clean validation data. We run the trained model on validation windows, collect the unsmoothed residuals $r_{k,\tau}$, and for each sensor $k$, we store the maximum observed value: $\theta_k \;=\; \max_{\tau \in \text{val}} r_{k,\tau}$. This baseline gives a conservative upper bound on normal reconstruction error, and at run-time, sensor $k$ is declared faulty at sequence $\tau$ when its smoothed residual exceeds its baseline (i.e., $\hat{r}_{k,\tau} > \theta_k$) (\pcirc{5}, \pcirc{6}).

Finally, to support multiple simultaneous failures, \textsc{Tureis} uses the same masking method as in training. After sensor $k$ is flagged at sequence $\tau$, all feature positions in $I_k$ are replaced with the \texttt{[MASK]} token in all subsequent sequences($t \ge \tau$). In fact, this isolation step prevents the faulty channel from affecting the reconstruction of other channels. Moreover, it allows additional sensors, whose errors were previously overshadowed, to surface in later windows. The process then continues from $\tau{+}1$ with the updated mask set, and any later sensor whose smoothed residual exceeds its baseline is handled in the same way. Because isolation is applied at the feature level and follows the original feature layout, the adjustment is immediate and does not require retraining.
\vspace{-3pt}
\section{Experimental Setup}
\label{ES}
\label{sec:setup}

\subsection{Evaluation Metrics}
\vspace{-3pt}
We use three complementary metric families: \textbf{\textit{precision}}, \textbf{\textit{recall}}, and \textbf{\textit{F1 score}} for detection and localization effectiveness, localization time for responsiveness, and computational overhead for edge feasibility. Together, these metrics evaluate how well \textsc{Tureis} meets its design goals. Precision is the fraction of true positives (i.e., actual positives detected correctly) among all detected positives, recall is the fraction of true positives among all actual positives, and F1 score is the harmonic mean of precision and recall. Thus, formally:
\[
\mathrm{precision} \;=\; \frac{\#\,\text{true positives}}{\#\,\text{predicted positives}},
\]
\vspace{1pt}
\[
\mathrm{recall} \;=\; \frac{\#\,\text{true positives}}{\#\,\text{actual positives}},
\]
\vspace{1pt}
\[
\mathrm{F1}\ \mathrm{score} \;=\; \frac{2 \times \mathrm{precision} \times \mathrm{recall}}{\mathrm{precision} + \mathrm{recall}}.
\]

Moreover, we define the \textbf{\textit{localization time}} as the time it takes to localize the faulty sensor. We also quantify \textbf{\textit{speed and edge efficiency}} using model size, parameter count, and training time. This also includes the end-to-end latency for each sequence, average CPU utilization, and peak RSS memory usage.

\subsection{Datasets}
\vspace{-3pt}
We use five publicly available, third-party smart-home datasets that have been used in the previous studies \cite{Choi2018Detecting,Borhani2022ThingsDND, Mallick2020DETECTIF, Jung2022Simultaneous, ElHady2020Sensor}. The suite includes one home from the Intelligent Systems Lab Amsterdam \cite{vankasteren2011activity}, \emph{HouseA}, and four homes from the Washington State University CASAS collection \cite{Cook2013CASAS}: \emph{HH102}, \emph{Tulum}, \emph{Atmo1}, and \emph{Tokyo}. These homes represent both single- and multi-resident settings, and include various types of sensors, such as motion, door/contact, appliance, and selected numeric channels. They also provide a wide range of daily routines and activity patterns. Table~\ref{tab:datasets} summarizes the number of residents and the types of sensors used in each dataset.

\begin{table}[t]
\centering
\scriptsize
\setlength{\tabcolsep}{2pt} 
\renewcommand{\arraystretch}{1.3} 
\caption{Summary of the datasets}
\vspace{-5pt}
\label{tab:datasets}
\begin{adjustbox}{width=\linewidth}
\begin{threeparttable}
\renewcommand\tabularxcolumn[1]{m{#1}} 
\begin{tabularx}{\linewidth}{l *{4}{>{\centering\arraybackslash}X}}
\toprule
\textbf{Dataset} & \textbf{Residents} & \textbf{Binary sensors} & \textbf{Numeric sensors} & \textbf{Activities} \\
\midrule
HouseA~\cite{vankasteren2011activity} & 1 & 14 & 0 & 16 \\
HH102~\cite{Cook2013CASAS}  & 1 & 33 & 79 & 30 \\
Tulum~\cite{Cook2013CASAS}  & 2 & 31 & 5 & 16 \\
Atmo1~\cite{Cook2013CASAS}  & 4 & 37 & 0 & 37 \\
Tokyo~\cite{Cook2013CASAS}  & 9 & 62 & 5 & --\tnote{a} \\
\bottomrule
\end{tabularx}
\begin{tablenotes}
\item[a] This dataset is not annotated
\vspace{-15pt}
\end{tablenotes}
\end{threeparttable}
\end{adjustbox}
\end{table}

\subsection{Baselines}
\vspace{-3pt}
To ensure a fair and repeatable evaluation, we compare \textsc{Tureis} with three representative baselines selected to cover key design tradeoffs. We outline the exact instantiations we evaluate and the controls we apply for parity, rather than restating each method (details in Section~\ref{sec:BackRW} and Table~\ref{tab:comparison_updated}).

\paragraph{DICE (context extraction + Markov checks)}
As DICE covers heterogeneous binary and numeric sensors, operates fully unlabeled with explicit context awareness, covers a wide range of data-fault types, and locates the exact faulty sensor at a low computational cost, we include it as our lightweight, non-ML baseline. It performs run-time correlation/transition checks for detection and localization after precomputing sensor correlations and transition probabilities~\cite{Choi2018Detecting}.  However, the original work targets single-failure, single-resident homes, and thus, we apply it to multi-resident homes to compare the results. Additionally, we set its threshold \texttt{numThre} to the number of faulty sensors in multi-failure scenarios. 

\paragraph{ThingsDND (context-aware BiLSTM)}
Since ThingsDND is demonstrably edge-friendly, supports a wide range of data-fault types and supports heterogeneous binary/numeric sensors, we include it as our compact, LSTM-based baseline. It is trained to model normal sequences from unlabeled streams using a BiLSTM over early-fused sensor features. Then, prediction errors were translated into per-sensor scores for detection and localization~\cite{Borhani2022ThingsDND}.  However, simultaneous multi-failure cases are not explicitly addressed in the original evaluation. Therefore, we set its rank-based thresholds to be equal to the distance between the highest score and the score at the rank equal to the number of faulty sensors, to ensure parity across multi-failure runs.

\paragraph{Anomaly Transformer (AT; reconstruction)}
 We use AT~\cite{Xu2022Anomaly} as our strong, adapted cross-domain baseline. The core hypothesis of AT is that anomalies exhibit a low Association Discrepancy, the KL-divergence between the model's learned series-association (i.e., the attention map) and a fixed, Gaussian-based prior-association. To create a fair and stable comparison on our datasets with discrete, event-based sensors and our binary, feature-engineered space, we retain the full AT architecture and its core minimax Association-Discrepancy loss objective and apply only two minimal, necessary adaptations. \textit{(i) Reconstruction objective:} The original AT is trained with a joint objective that includes an L2 reconstruction loss, which is suitable for continuous, unbounded data. We replace this component with a binary cross-entropy objective, which is the appropriate loss function for reconstructing our binary feature vectors. \textit{(ii) Inference score:} The original scoring function applies a batch-wise softmax over sequences to weight the reconstruction error, which is unstable for comparison, as its output is normalized based on the number of sequences being scored. We therefore use a simple per-sequence weighting scheme that increases the weight as Association Discrepancy decreases, preserving the original emphasis of AT on low-discrepancy sequences.

\subsection{Evaluation Protocol}
\vspace{-3pt}
For each dataset listed in Table~\ref{tab:datasets}, we reserve the first 600 hours for offline preparation (i.e., feature extraction and model training) and the next 180 hours for run-time execution.  No failures are injected during the 600-hour block, which consists of the first 500 hours for training and the following 100 hours for clean validation. The 180-hour evaluation horizon is partitioned into 30 segments of 6 hours each. Then, we duplicate each segment: the duplicate receives injected failures to evaluate detection, per-sensor localization, and localization time, while the original stays clean and measures nominal behavior. We uniformly and randomly sample a sensor, a fault type (as defined in Section~\ref{sec:BackRW}), and an insertion time in each injected segment. The above-mentioned split is intentional, and the 180-hour window is just the remaining time after setting aside 600 hours for offline preparation to ensure that training, validation, and testing data do not overlap. Moreover, a 6-hour segment is both short enough to yield many independent trials per home and long enough to capture diverse activity chains, observe failures, and subsequent stabilization (hundreds of 5-minute sequences per segment). All speed and edge efficiency measurements of \textsc{Tureis} are based on our Raspberry~Pi~5 (quad-core 2.4\,GHz Cortex-A76, 8\,GB RAM).

\section{Evaluation}
\label{Eval}
\label{sec:evaluation}

\subsection{Single-Failure Results}
\vspace{-3pt}
\subsubsection*{\textbf{Failure Detection}}

\begin{figure*}[t]
    \centering
    \includegraphics[width=1\textwidth,trim=5 9 7 7,clip]{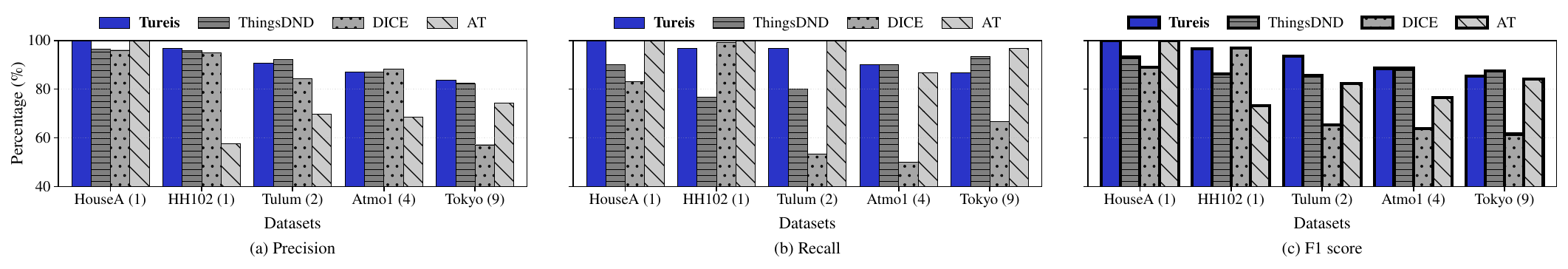}
    \vspace{-20pt}
    \caption{\textbf{Single-failure detection:} precision, recall, and F1 score for \textsc{Tureis}, ThingsDND, DICE, and AT on five smart homes (the number of residents is shown in parentheses). F1 score (i.e., the harmonic mean of precision and recall) is shown with a bold outline to emphasize its role as the primary summary metric.}
    \vspace{-7pt}
    \label{single-det-prf}
\end{figure*}

For failure detection, evaluation metrics should indicate how successfully the proposed method detects failures. Therefore, precision in this context is the fraction of segments correctly detected as faulty among all segments detected as faulty. Moreover, recall is the fraction of the segments correctly detected as faulty among all actual faulty segments. Precision, recall, and F1 scores for single-failure detection are reported across smart home datasets in Figure~\ref{single-det-prf} (dataset name followed by resident count in the figure). \textsc{Tureis} achieves the best precision (Figure~\ref{single-det-prf}(a)) on three homes and the best recall (Figure~\ref{single-det-prf}(b)) on two, and thus, delivers the top F1 score (Figure~\ref{single-det-prf}(c)) on four of five homes (\emph{HouseA}, \emph{HH102}, \emph{Tulum}, \emph{Atmo1}). Moreover, it is within a narrow margin of the best method (i.e., ThingsDND) on \emph{Tokyo} (9 residents). Therefore, we can conclude that ThingsDND remains competitive in terms of single-failure detection F1 score on multi-resident homes, while DICE and AT fall behind as the number of residents increases.

The results are in line with \textsc{Tureis} design: first, consistently high precision and recall are achieved by combining bit-level, early-fused features with per-sensor validation calibration to reduce false positives and increase true positives across heterogeneous channels. Second, the model is forced to use cross-sensor and temporal correlation in sensor-wise masked reconstruction with bidirectional self-attention, which enables it to recover weak, context-dependent evidence. Moreover, EWMA smoothing further suppresses one-off spikes, lowering false positive rates. Together, these effects shift the precision-recall curves upward, producing the best or tied-for-best F1 score across homes. However, as we move from \emph{HouseA} (1) to \emph{Tokyo} (9), an increase in the number of residents leads to more simultaneous resident activities, which dilute sensor correlations and makes it difficult to detect the failures. As a result, all methods become harder pressed.

\subsubsection*{\textbf{Faulty Sensor Localization}}
\begin{figure*}[t]
    \centering
    \includegraphics[width=1\textwidth,trim=5 9 7 7,clip]{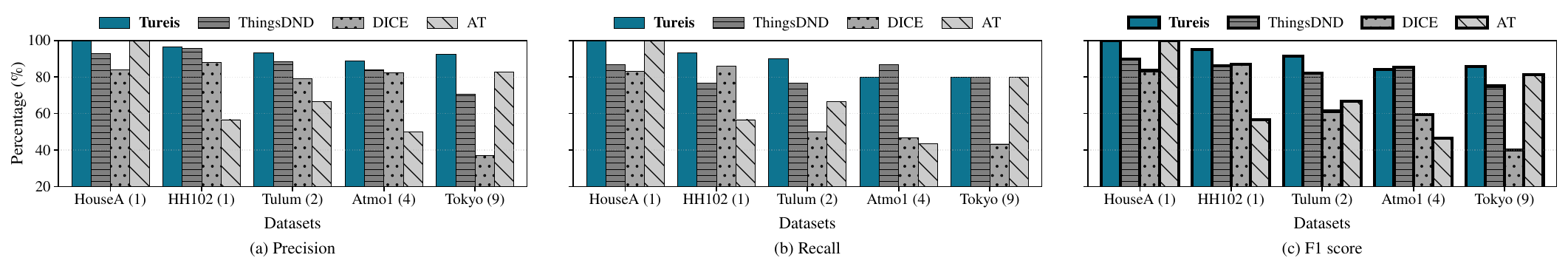}
    \vspace{-20pt}
    \caption{\textbf{Single-failure localization:} precision, recall, and F1 score across the five smart homes (resident counts as in labels). F1 score is bolded as the primary summary metric.}
    \vspace{-17pt}
    \label{single-diag-prf}
\end{figure*}

For faulty-sensor localization, metrics indicate how successfully a method locates the faulty sensor. Thus, precision is the fraction of correctly located sensors among all located sensors, and recall is the fraction of correctly located sensors among all actual faulty sensors. As shown in Figure~\ref{single-diag-prf}(a–c), \textsc{Tureis} achieves the highest precision on all five homes and the highest recall on four of five, resulting in the top F1 score on \emph{HouseA}, \emph{HH102}, \emph{Tulum}, and \emph{Tokyo}. ThingsDND achieves a marginally higher F1 score and a slightly higher recall on \emph{Atmo1}, which only contains binary sensors and lacks numeric streams. Our analysis suggests that some of these binary channels exhibit high clean-time variability, which inflates their per-sensor max baselines \(\theta_k\) and makes it harder for our method to trip true failures (i.e., lowering recall). In contrast, ThingsDND thresholding, which is not calibrated per sensor in the same way and is less sensitive to such per sensor variability, can maintain a slightly higher F1 score on Atmo1. Overall, ThingsDND is still competitive for single-failure localization on multi-resident homes, while AT and DICE have lower F1 score due to their relatively low recall. 

While the AT baseline is a strong competitor that uses the full association-discrepancy objective, \textsc{Tureis} still shows a consistent advantage. This highlights the benefits of \textsc{Tureis}'s sensor-wise masked-reconstruction objective, which is domain-specific and explicitly designed to learn sensor relationships, versus AT's more general-purpose objective. Furthermore, \textsc{Tureis}'s simple residual scoring and per-sensor calibration proves more robust for localization than AT's combined scoring metric.
Furthermore, localization is stricter than detection, which amplifies differences among methods and typically yields lower results. Nevertheless, consistent with detection, \textsc{Tureis}'s high F1 score is a result of its early-fused features, masked reconstruction, and per-sensor calibration.

\subsubsection*{\textbf{Localization time}}

\begin{figure}[t]
    \centering
    \includegraphics[width=1\linewidth,trim=5 5 5 5,clip]{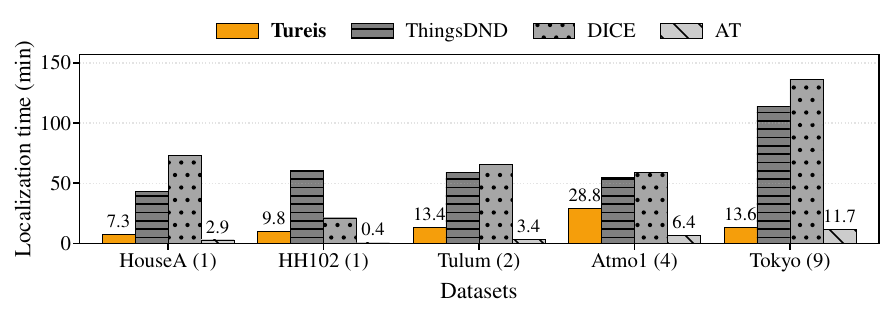}
    \vspace{-22pt}
    \caption{\textbf{Single-failure localization time:} Mean delay (minutes) from ground-truth failure start to the first correct localization for \textsc{Tureis}, ThingsDND, DICE, and AT across five homes.}
    \label{single-latency}
    \vspace{-17pt}
\end{figure}

\begin{figure*}[t]
    \centering
    \includegraphics[width=1\textwidth,trim=5 9 7 7,clip]{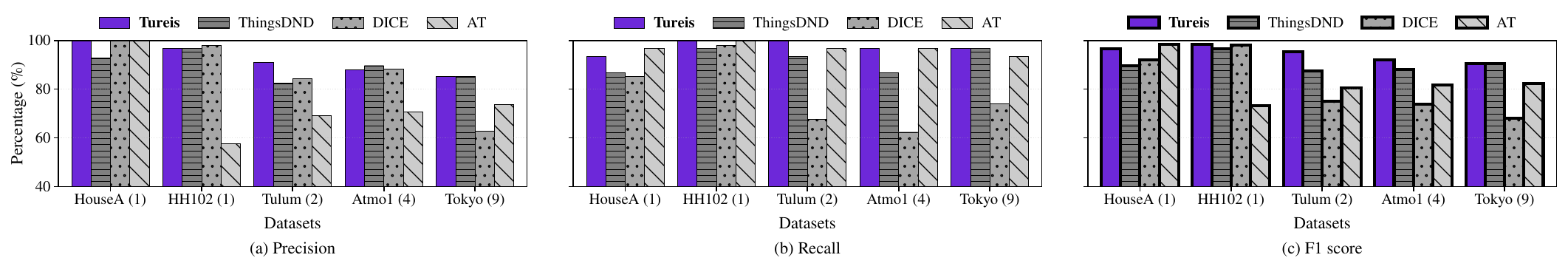}
    \vspace{-20pt}
    \caption{\textbf{Multi-failure detection:} precision, recall, and F1 score across the five smart homes (resident counts shown in labels). Up to five simultaneous failures are injected, and F1 score is bolded as the primary summary metric.}
    \vspace{-10pt}
    \label{multi-det-prf}
\end{figure*}

\begin{figure*}[t]
    \centering
    \includegraphics[width=1\textwidth,trim=5 9 7 7,clip]{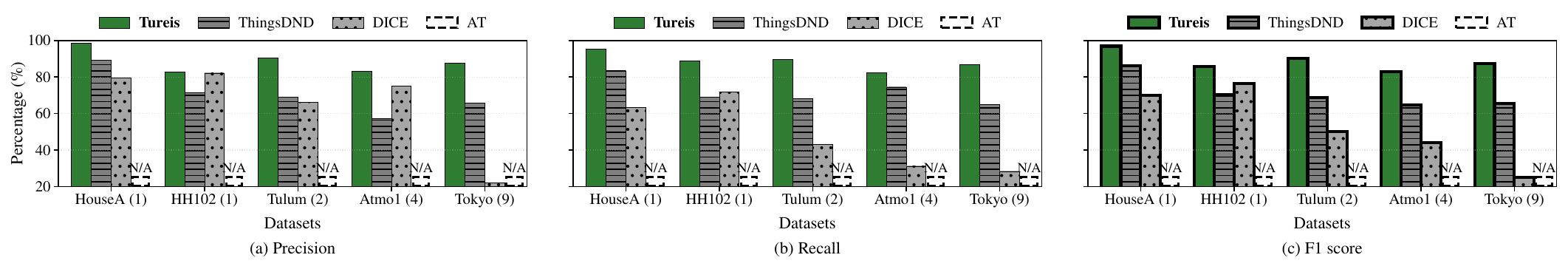}
    \vspace{-20pt}
   \caption{\textbf{Multi-failure localization:} precision, recall, and F1 score across five smart homes (resident counts in labels). Up to five simultaneous failures are injected, and F1 score is bolded as the primary summary metric. Moreover, AT is not applicable to multi-failure localization and is shown as N/A.}
   \vspace{-15pt}
    \label{multi-diag-prf}
\end{figure*}

We measure localization time as the mean delay (minutes) from the ground-truth failure start to the first correct localization decision. As demonstrated in Figure~\ref{single-latency}, \textsc{Tureis} is significantly faster than ThingsDND and DICE on all datasets (often by \(3{\times}\)–\(10{\times}\)), and remains within a small constant factor of AT. AT attains shorter localization time because its anomaly score is highly sensitive to abrupt changes in sensor behavior, so its scores cross the decision threshold earlier. However, this aggressive sensitivity also makes it more prone to benign fluctuations, contributing to its lower precision. In contrast, \textsc{Tureis} generates confident per-sensor residuals earlier than ThingsDND's BiLSTM which relies on sequential prediction, and DICE's Markov checks, which require multiple transition violations to accumulate evidence. To be more specific, the consistently low mean delays are explained by EWMA-smoothed residuals over brief 5-minute windows with a 1-minute stride, which offer timely evidence without over-triggering. Additionally, \textsc{Tureis} employs parallel bidirectional self-attention and sensor-wise masked reconstruction to fuse cross-sensor correlations within each window. In fact, localization latency trades off precision, recall, and F1 score. Using shorter windows or strides reduces delay but lessens context, which increases misses and false positives. However, with our near-Pareto configuration, \textsc{Tureis} effectively balances mean localization time with quality, achieving low mean localization times alongside high precision, recall, and F1 scores.

\subsection{Multi-Failure Results}
\vspace{-3pt}
For the multi-failure setting, the only modification is injection cardinality, and everything else is the same as in the single-failure case. Specifically, each segment contains up to five simultaneous injected failures. We do this by sampling the number of simultaneous failures from a Poisson distribution with mean \(\lambda{=}3\) and clipping it to \([1,5]\). This approach provides varied but realistic multi-failure loads without overemphasizing extremes by concentrating probability around three failures. 

\subsubsection*{\textbf{Failure Detection}}
Figure~\ref{multi-det-prf}(a–c) reports multi-failure detection, and demonstrates that \textsc{Tureis} attains the best F1 score on all five homes. Moreover, in terms of precision, \textsc{Tureis} is highest on three homes, and in recall, it leads on four of five datasets. As in the single-failure case, the benefit of \textsc{Tureis} is derived from the same synergistic design decisions: bit-level, early-fused features with per-sensor validation calibration, sensor-wise masked reconstruction with bidirectional self-attention, and EWMA smoothing. When combined, these factors produce the best F1 across homes by maintaining a superior precision–recall balance in the multi-failure setting. Moreover, compared to the single-failure setting, detection is generally easier here because multiple simultaneous failures increase the evidence of abnormality within the segment, and thus, all methods improve.

\subsubsection*{\textbf{Faulty Sensor Localization}}

Figure~\ref{multi-diag-prf}(a–c) reports multi-failure localization and demonstrates that \textsc{Tureis} delivers the strongest performance overall and clearly excels, attaining the top F1 across all datasets, with the best precision and leading recall on all datasets. Two design decisions drive multi-failure localization gains of \textsc{Tureis}. First, sensor-wise masked reconstruction with bidirectional self-attention enforces temporal and cross-sensor reasoning, recovering subtle, context-dependent evidence even under simultaneous failures. Second, the masked-reconstruction training objective is naturally extended by iterative masking and isolation at run-time. Specifically, after a sensor is located as faulty, \textsc{Tureis} masks its channel moving forward to avoid interfering with other channels and to allow more failures to surface one after the other.

On the other hand, ThingsDND BiLSTM aggregates evidence sequentially and tends to smear overlapping failures across time for multi-resident homes with dilute sensor correlations, which can blur per-sensor localization, resulting in decreased both recall and precision.. Furthermore, DICE relies on Markov transition checks, which can reveal a significant number of potentially faulty sensors when several failures happen at once. Additionally, AT lacks sensor-wise masking and iterative isolation, and it combines reconstruction error and association discrepancy into a single score. As a result, when one channel fails, its large residual takes over the score, and since it lacks a mechanism to suppress that channel, other failures remain hidden. In other words, AT functions similarly to a single arg-max selector in the absence of a process to expose several faulty sensors one after the other. Therefore, multi-failure localization is beyond the capabilities of AT.

\begin{table*}[htbp]
\centering
\scriptsize
\caption{Speed and edge efficiency of \textsc{Tureis} across all datasets. Offline (training) results use 500~hours of clean data, and run-time (inference) results are aggregated over the 180-hour evaluation horizon.}
\vspace{-10pt}
\label{tab:edge-efficiency}
\begin{adjustbox}{width=\linewidth}
\begin{tabular}{lcccccccccc}
\toprule
 & \multicolumn{4}{c}{Offline (training)} & \multicolumn{5}{c}{Run-time (inference)} \\
\cmidrule(lr){2-5} \cmidrule(lr){6-10}
\textbf{Dataset}
& \textbf{Size [MB]}
& \textbf{Params [M]}
& \textbf{Time [min]}
& \textbf{Peak RAM [MB]}
& \textbf{Lat (clean) [ms]}
& \textbf{Lat (single) [ms]}
& \textbf{Lat (multi) [ms]}
& \textbf{CPU [\%]}
& \textbf{RAM [MB]} \\
\midrule
HouseA & 0.283 & 0.037 & 65.0 & 451 & 1.401 & 1.455 & 3.311 & 150.6 & 415 \\
HH102  & 0.484 & 0.090 & 74.5 & 866 & 1.548 & 1.582 & 5.217 & 168.2 & 516 \\
Tulum  & 0.309 & 0.044 & 68.3 & 436 & 1.442 & 1.475 & 3.776 & 123.8 & 391 \\
Atmo1  & 0.305 & 0.043 & 67.3 & 478 & 1.441 & 1.478 & 3.837 & 147.1 & 434 \\
Tokyo  & 0.339 & 0.052 & 67.9 & 528 & 1.466 & 1.5 & 4.445 & 124.4 & 445 \\
\midrule
\textbf{Avg}    & \textbf{0.344} & \textbf{0.053} & \textbf{68.6} & \textbf{552} & \textbf{1.46} & \textbf{1.498} & \textbf{4.117} & \textbf{142.8} & \textbf{440} \\
\bottomrule
\end{tabular}
\end{adjustbox}
\vspace{-10pt}
\end{table*}

\subsubsection*{\textbf{Localization Time}}
\begin{figure}[t]
    \centering
    \includegraphics[width=1\linewidth,trim=5 5 5 5,clip]{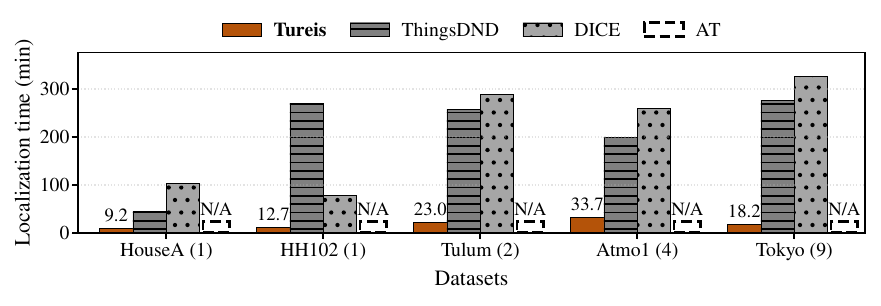}
    \vspace{-20pt}
    \caption{\textbf{Multi-failure localization time:} mean delay (minutes) from failure onset to the first correct localization. Up to five simultaneous failures are injected. N/A indicates that AT is not applicable to multi-failure localization.}
    \vspace{-15pt}
    \label{multi-latency}
\end{figure}
As shown in Figure~\ref{multi-latency}, \textsc{Tureis} yields the shortest mean delays on every home, while AT is not applicable to per-sensor multi-failure localization (marked N/A). Compared to ThingsDND and DICE, \textsc{Tureis} locates faulty sensors faster because short sequences (\(L{=}5\), \(stride{=}1\)), EWMA smoothing surface evidence promptly, and the isolate-and-continue loop suppresses already-flagged channels so subsequent faults emerge without interference.  In comparison to single-failure cases, the mean localization time of \textsc{Tureis} grows only modestly across homes, whereas ThingsDND and DICE slow down substantially under multi-failure settings. This is because ThingsDND declares a failure only when the score distance between the top-ranked sensor and the fifth-ranked sensor exceeds a validation-derived threshold. However, BiLSTM prediction errors inflate scores for many sensors during simultaneous failures, making the distance widen slowly. Thus, more timesteps are required before the distance crosses the threshold. Moreover, the Markov transition checks of DICE generate large candidate sets when multiple failures occur. Therefore, it needs to shrink these sets via repeated transition checks, and thus, the set decays slowly, which requires many iterations to converge to the true sensors and thereby increases localization time.

\subsection{Per-Fault-Type Analysis}
\vspace{-3pt}
\begin{figure}[t]
    \centering
    \includegraphics[width=1\linewidth,trim=5 5 5 7,clip]{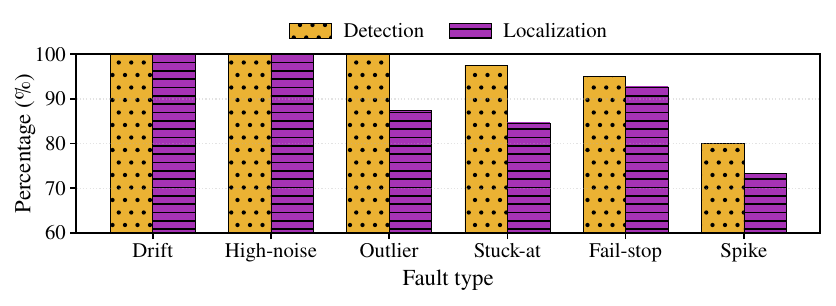}
    \vspace{-20pt}
    \caption{\textbf{Per-fault-type results:} percentage of injected failures that are detected and located for six fault types. Results are aggregated over all five smart homes.}
    \vspace{-15pt}
    \label{per-fault-single}
\end{figure}

Figure~\ref{per-fault-single} reports the percentage of injected failures that were detected and localized per fault type. Drift and high-noise reach 100\% for both detection and localization across homes, consistent with the masked-reconstruction design of \textsc{Tureis}: on numeric channels, jumpy and burst dynamic bits amplify intervals that are dominated variance and trend, producing sustained residuals that make attribution to the faulty sensor straightforward. By contrast, outliers perturb only a few interval vectors, yielding a smaller residual footprint and thus slightly lower results than drift and high-noise. Stuck-at is persistent in value but can be less separable when the held value lies within the normal operating range of the sensor. Such flat segments remain temporally consistent, and masked reconstruction may shift evidence toward correlated neighbors. Conversely, fail-stop presents the opposite signature, and the channel goes completely silent. This silence can likewise preserve temporal consistency and make threshold crossings difficult. Finally, spike is the most challenging among the six fault types. Although spike produces significant changes, it can induces a quasi-consistent pattern (i.e., a sharp transition followed by stable readings), which preserves the temporal correlation. Thus, it yields a narrower residual footprint than the other fault types.

\subsection{Speed and Edge Efficiency}
\vspace{-3pt}
Smart home systems rely on IoT components, including sensors, actuators, aggregators, and communication channels, that together generate continuous data streams\cite{Gubbi2013Internet, Voas2016Networks}. Although cloud resources offer additional computation, sending large volumes of sensor data to remote servers incurs bandwidth overhead and latency, which is undesirable for real-time applications\cite{Kong2023Edge, Elgamal2018Distributed}. Therefore, sensor failure detection must be lightweight and directly deployable on edge devices, such as aggregators, including the home gateway (i.e., the last-level aggregator). Guided by these constraints, our work designs an edge-friendly method that operates locally without relying on cloud resources. Table~\ref{tab:edge-efficiency} summarizes both the one-time offline training cost and the run-time resource footprint of \textsc{Tureis} on our Raspberry~Pi~5 board.

\subsubsection*{\textbf{Offline Training Cost}}
For each smart home, we train \textsc{Tureis} on 500 hours of clean feature streams and then we measure the model size, parameter count, training time, and peak training memory directly on the Raspberry~Pi~5 (offline columns in Table~\ref{tab:edge-efficiency}). This experiment demonstrates that the entire training process of \textsc{Tureis} can be easily completed on a low-cost home gateway. The encoder stays sub-megabyte, with only tens of thousands of parameters, and a full training run per home is completed in about an hour while keeping peak memory well within the capacity of the device. Thus, purely on-device training is feasible, which is appealing for privacy-preserving deployments and homes without cloud connectivity. Moreover, as expected, offline training cost increases with the dimensionality of the bit-level feature space, which in turn is determined by the number of sensors in each home.

\subsubsection*{\textbf{Run-time Edge Efficiency}}
To evaluate the run-time cost of continuous failure detection and localization on the Raspberry~Pi~5, we run \textsc{Tureis} over the full 180-hour evaluation horizon and measure latency (reported as computation time per minute of data, reflecting the processing needed per time interval), CPU utilization, and peak RAM under three regimes: clean streams, single-failure runs, and multi-failure runs (run-time inference columns in Table~\ref{tab:edge-efficiency}). Across all homes, clean and single-failure streams require only around \(1\)–\(2\)~ms of computation per minute of smart-home data. Actually, single-failure latency is close to the clean case because failures are rare, so most windows are processed exactly as in clean streams. 

In contrast, multi-failure streams introduce more frequent isolate-and-continue logic, but even then, the per-minute latency stays within roughly \(3\)–\(6\)~ms. Thus, the Raspberry~Pi spends only a tiny portion of each real-time minute on \textsc{Tureis}, leaving ample headroom for other services. Moreover, CPU utilization stays moderate, corresponding to roughly one to two active cores on the Raspberry~Pi. Peak run-time memory also stays well below 520~MB in all cases, comfortably fitting within the capacity of a commodity 8~GB device. Together with the offline results, these measurements confirm that, despite using a Transformer, \textsc{Tureis} remains lightweight and edge-friendly, due to its bit-level feature encoding and compact encoder-only architecture.

\vspace{-3pt}
\section{Conclusions}
\label{Conc}
\label{sec:conclusion}
This paper presented \textsc{Tureis}, a fully self-supervised, context-aware method for failure detection and faulty-sensor localization in smart homes, designed for deployment on resource-constrained edge devices. Offline, the method couples bit-level, context-aware feature extraction with a lightweight, short-horizon Transformer encoder trained via sensor-wise masked reconstruction. The feature extraction aggregates raw heterogeneous sensor streams into compact bit-level features, and the encoder captures short-horizon spatiotemporal correlations from these unlabeled features. At run-time, \textsc{Tureis} converts reconstruction residuals into per-sensor failure evidence. Once a sensor trips, it suppresses the feature stream of that sensor, eliminating interference and allowing additional failures to be localized. To our knowledge, \textsc{Tureis} is the first method that simultaneously handles multi-failure, multi-resident smart homes, covers diverse fault types on heterogeneous sensors, operates fully unlabeled, and remains deployable on edge devices. Our evaluation across five third-party datasets shows that \textsc{Tureis} consistently improves single- and multi-failure detection and localization F1 scores over three strong state-of-the-art baselines. It localizes faulty sensors within a few minutes of failure onset, while keeping the encoder in the sub-megabyte range. On a Raspberry~Pi~5, it processes each minute of data in only a few milliseconds, with low CPU utilization and a compact memory footprint.



\bibliographystyle{IEEEtranS}
\bibliography{tureis}

\end{document}